\documentclass[fleqn,10pt]{wlscirep}
\usepackage[utf8]{inputenc}
\usepackage[T1]{fontenc}

\title{Customizable wave tailoring materials enabled by nonlinear bilevel inverse design}

\author[1,+]{Brianna MacNider}
\author[1,+]{Haning Xiu}
\author[1]{Kai Qian}
\author[1]{Ian Frankel}
\author[2,3]{Hyunsun Alicia Kim}
\author[1,3,*]{Nicholas Boechler}
\affil[1]{Department of Mechanical and Aerospace Engineering, University of California, San Diego, La Jolla, CA 92093,US}
\affil[2]{Department of Structural Engineering, University of California, San Diego, La Jolla, CA 92093,US}
\affil[3]{Program in Materials Science and Engineering, University of California, San Diego, La Jolla, CA 92093, US}

\affil[*]{nboechler@ucsd.edu}

\affil[+]{these authors contributed equally to this work}

\begin{abstract}
Passive transformation of waves via nonlinear systems is ubiquitous in settings ranging from acoustics to optics and electromagnetics. Passivity is of particular importance for responding rapidly to stimuli and nonlinearity enormously expands signal transformability compared to linear systems due to the breaking of superposition. It is well known that different types of nonlinearity yield vastly different effects on propagating signals, which raises the question of ``what precise nonlinearity is the best for a given wave tailoring application?'' Considering a one-dimensional spring-mass chain as a testbed, we couple the shape optimization of structures for tailored nonlinear constitutive responses with reduced-order nonlinear dynamical inverse design. Using minimization of peak kinetic energy transmission from impact as a case study, we identify ideal nonlinear constitutive responses and the geometries needed to achieve them. As part of this, we show the large sensitivity of this metric to small changes in nonlinearity, and thus the need for high precision, free-form nonlinearity tailoring. We validate our predictions using impact experiments in a chain of nonlinear springs and masses. This work sets the foundation for broader passive nonlinear mechanical wave tailoring material design, with applications to computing, signal processing, shock mitigation, and autonomous materials.\\

\textbf{keywords:} nonlinear dynamics, nonlinear mechanics, impact mitigation, nonlinear waves, inverse design
\end{abstract}

\begin{document}

\flushbottom
\maketitle

\thispagestyle{empty}

\section*{Introduction}

\noindent The passive transformation of waves via nonlinear material response is widely used in physical settings ranging from acoustics \cite{patil2022review} to optics \cite{boydbook,kauranen2012nonlinear,litchinitser_nonlinear_2018} and electromagnetics \cite{KivsharNonlinearMeta,fairbanks2020review}. Applications include areas such as efficient information transfer \cite{wang2020advances, smirnova2020nonlinear}, computing and logic \cite{marcucci2020theory,yasuda2021mechanical}, energy conversion \cite{dmitriev2013handbook}, imaging \cite{wang2023image}, encryption \cite{zhang2014symmetric}, impact and vibration mitigation \cite{nesterenko2013dynamics,fancher2023dependence}, and rapid shape change \cite{jin2020guided}. Within these contexts, in contrast to active control, passivity is of particular importance for responding fast to stimuli, and nonlinearity enormously expands signal transformability compared to linear systems due to the breaking of superposition. Indeed, it is well known that different types of nonlinearity yield vastly and qualitatively different effects on propagating signals \cite{scott2006encyclopedia}, which raises the question of ``what precise nonlinearity is the best for a given wave tailoring application?'' This question has largely remained in the regime of simulation and theory, as, until recently, it has not been possible to freely realize any optimal nonlinear constitutive law in practice. The field of mechanics has come furthest towards this goal, by introducing complex, sub-wavelength, geometric motifs to create ``mesostructured'' nonlinear materials \cite{patil2022review}, however any tunability seen has been limited around a handful of known nonlinear mechanisms. For instance, broad classes of nonlinearity that have seen tailorability for wave manipulation include contact nonlinearities \cite{porter_granular_2015}, tensegrity structures \cite{fraternali_multiscale_2014}, and bistable beam arrays \cite{meaud_tuning_2017}, among others.

Recent progress has enabled a, thus far unique-to-mechanics, capacity to create materials with on-demand quasi-static nonlinear properties via shape and structural optimization \cite{li_design_2021,li_digital_2022,QuasiOptArxivPlaceholder,deng_inverse_2022,zheng_unifying_2023,brown_deep_2023,ha2023rapid,bastek2023inverse}. This has included several approaches, including gradient based topology optimization in pursuit of tailoring the entirety of a nonlinear force-displacement curve \cite{ li_design_2021,li_digital_2022,QuasiOptArxivPlaceholder} as well as the incorporation of machine learning (ML) algorithms in an attempt to traverse the design space and speed up predictions of mechanical behavior \cite{zheng_deep_2023,deng_inverse_2022,brown_deep_2023,ha2023rapid,bastek2023inverse}. However, such methods alone cannot identify material designs for optimal system-level nonlinear wave tailoring performance. Prior studies of optimal nonlinear dynamic material behavior have tailored heterogeneity with fixed nonlinearity \cite{fraternali2009optimal}, or dynamic behavior where the characteristic wavelengths are on par with or greater than the system size (and thus the response is not ``wave-dominated'') and the tailoring was confined to broad metrics like ``area under the curve'' \cite{deng_topology_2020,chen_design_2018} or ``plateau-like'' behavior \cite{https://doi.org/10.1002/advs.202204977}. The role of waves is of particular importance, as allowing for spatiotemporal evolution in nonlinear systems leads to unique emergent phenomena such as solitons \cite{dauxois2006physics}. The role of precisely engineered nonlinearities is further important for wave propagation in that seemingly subtle differences in nonlinearity yield \emph{qualitatively} different dynamical behavior. For one example, consider a material with polynomial nonlinearity and all positive coefficients, resulting in a ``stiffening'' nonlinearity: just small changes in the ratio of coefficients dictate whether or not the system experiences modulational instability \cite{boechler2010discrete,MIhu}, or the difference of a quadratic versus cubic perturbation on a linear stiffness results in qualitatively different waveforms and wave mixing behavior \cite{boydbook}. Connecting the inverse design of nonlinear wave response to the quasi-static design of nonlinear constitutive response induced by mesostructure geometry is a significant, and hitherto unsurmounted challenge. If trying to directly extend quasi-static geometric design algorithms based on finite element method (FEM) simulation \cite{li_design_2021,li_digital_2022,QuasiOptArxivPlaceholder,deng_inverse_2022,zheng_unifying_2023,brown_deep_2023,ha2023rapid,bastek2023inverse} the challenge becomes evident, in that one would need to take the same design variables, copy the geometry over many unit cells, and simulate the entire system in time at high temporal resolution (due to nonlinear generation of high frequency content), and wrap that in an automated design loop---resulting in a task of extreme computational expense. 

\begin{figure}[ht!]
\centering
\includegraphics[width=17.5cm]{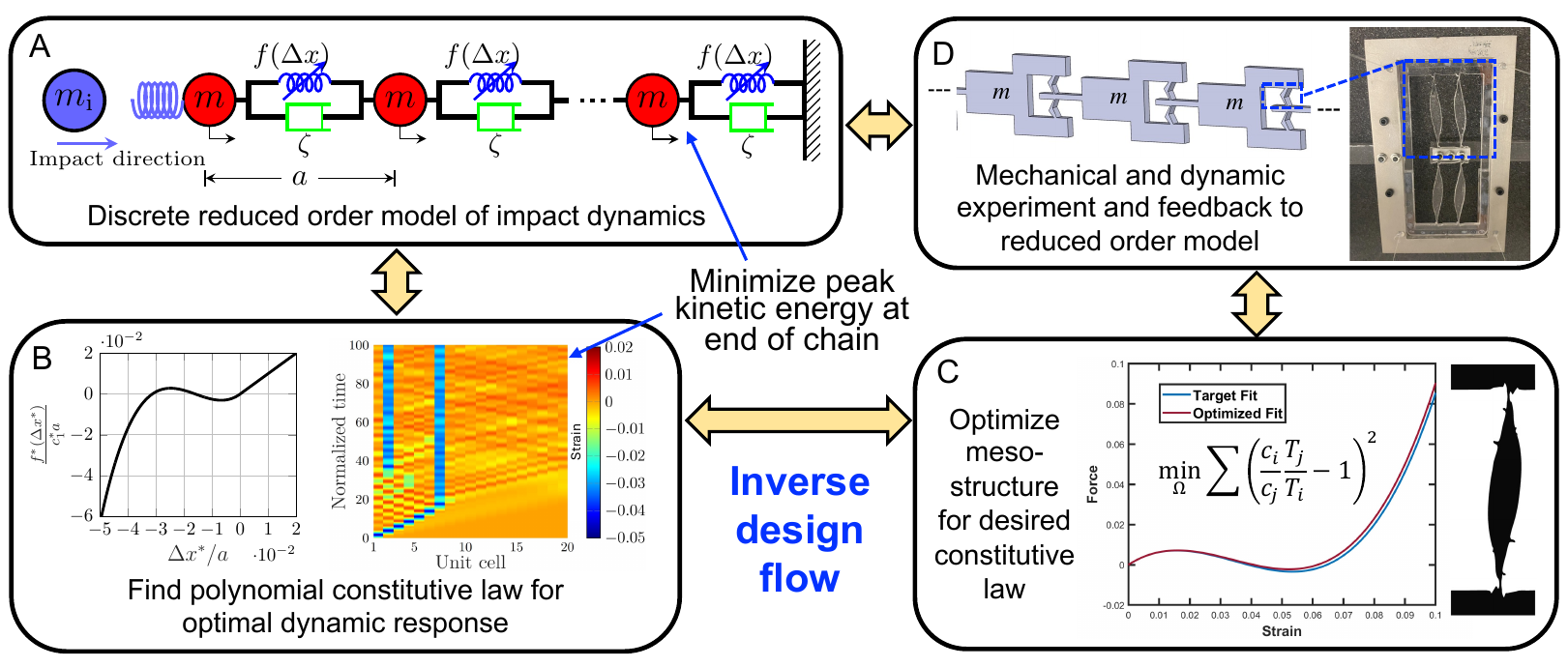}
\caption{{\bf Overview of the bilevel design flow}. (A) Discrete element model (DEM) simulation of the entire system dynamics. (B) Identification of the optimal nonlinear constitutive law. (C) Shape optimization of a mesostructure to match the identified nonlinear constitutive law. (D) Mechanical and dynamic experimental system characterization.}\label{inverse_design_chart_flow}
\end{figure}

In this work, we introduce a method to create customizable wave tailoring materials via nonlinear bilevel inverse design. Namely, we optimize for the emergent dynamic response of a mesostructured material in the form of a one-dimensional (1D) spring-mass chain. To do this, we use a reduced order, discrete element model (DEM) simulation to identify an optimal nonlinear constitutive law for the given performance metric, and couple this to a unit-cell-scale, geometrically-nonlinear, free-form, shape optimization algorithm which designs a physical system that achieves the nonlinear constitutive property identified by the DEM (outlined in Fig.~\ref{inverse_design_chart_flow}). We note that the 1D spring-mass chain is a prototypical nonlinear system that has classically formed the foundation for extensions into higher dimensions and continua \cite{porter2009fermi,dauxois2006physics,scott2006encyclopedia}, and, as part of our work herein, we illustrate extensions of our optimized unit-cells to two- and three-dimensional analogs (see Supplementary Information Note 1). Further, we demonstrate the clear need for high precision nonlinear response design, wherein the system dynamic performance is shown to be highly sensitive to the nonlinear coefficients of the constitutive law. Herein, we choose the maximum kinetic energy transmitted to the boundary of the system in response to an impact event, normalized by that of an otherwise-identical linear system, as our performance metric. The comparison between the linear and nonlinear response is particularly important, as it isolates the role of nonlinearity from other wave manipulating effects such as dispersion and dissipation. We demonstrate the full inverse design of this superior impact mitigation system, from identification of an ideal nonlinearity, to the design of the unit cell geometry, and experimental validation of the performance. An important feature to note, unlike some prior computational quasi-static nonlinear mechanical design strategies \cite{li_digital_2022,ha_rapid_2023,deng_inverse_2022,zheng_unifying_2023}, we do not use simplified or reduced order models for our underlying mesostructure design, which enables a broader design space and access to highly precise tailoring of nonlinear responses \cite{QuasiOptArxivPlaceholder}. While we demonstrate our bilevel inverse design for the case of impact mitigation, we expect such an approach for the previously specified nonlinear wave transformation applications \cite{patil2022review,wang2020advances,smirnova2020nonlinear,marcucci2020theory,yasuda2021mechanical,dmitriev2013handbook,wang2023image,zhang2014symmetric} within the acoustic, phononic, and mechanical wave settings.

\section*{Results}

\subsection*{Reduced order discrete element model}

The DEM (Fig. \ref{inverse_design_chart_flow}A) simulates the impact of a rigid ``impactor'' block onto a chain with $N$ unit cells and $a$ unit cell length. We represent the chain as lumped masses of mass $m$, interconnected by massless nonlinear springs and inter-site linear dampers in parallel, emulating the behavior of a viscoelastic material. The nonlinear spring consists of a linear spring behavior in tension and an up-to-third-order polynomial in compression, where the linear stiffness remains fixed at $c_1^\ast$ in both tension and compression (as shown in Fig.~\ref{inverse_design_chart_flow}B). This results in non-dimensionalized equations of motion for our chain
\begin{equation}\label{dem_nondimension}
   \ddot{x}_{i}-(x_{i+1}-x_{i})+c_2(x_{i+1}-x_{i})^2-c_3(x_{i+1}-x_{i})^3
   +(x_{i}-x_{i-1})-c_2(x_{i}-x_{i-1})^2+c_3(x_{i}-x_{i-1})^3+2\zeta(-\dot{x}_{i+1}+2\dot{x}_{i}-\dot{x}_{i-1})=0,
\end{equation}

\noindent where dimensionless stiffnesses $c_2$ and $c_3$ are zero when that spring is in tension, $\zeta$ is the inter-site damping ratio, $x_{i}$ is the dimensionless displacement of the $i$th particle from its rest position, overdots represent the derivative with respect to nondimensional time, and all variables and parameters are normalized by combinations of $a$, $m$, and/or $c_1^\ast$, as is described in Supplementary Information Note 2. We note that while the masses are illustrated as $m$ in Fig.~\ref{inverse_design_chart_flow}A, the nondimensional mass of each particle remains one as in Eq.~(\ref{dem_nondimension}). A variable mass and velocity rigid impactor is applied on the left, and a fixed boundary on the right. The model also incorporates a contact spring designed to facilitate the smooth contact and controlled release of the impactor during initial impact and rebound, respectively. The simulated dynamical response is acquired through the numerical integration of Eq.~(\ref{dem_nondimension}) via a Runge-Kutta algorithm. The non-dimensionalization of all variables and full details concerning the equations of motion are described in Supplementary Information Note 2.

\subsection*{Optimization of nonlinear constitutive law based on dynamical response}
 
\begin{figure}[ht!]
\centering
\includegraphics[width=17.5cm]{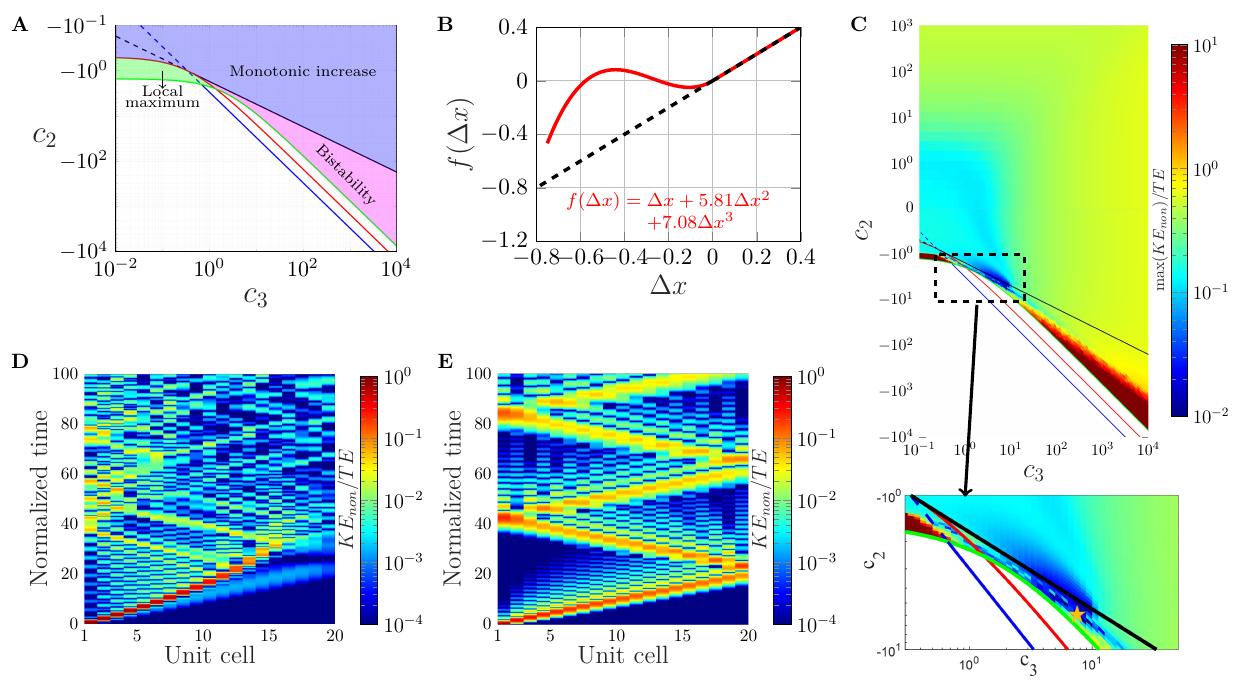}
\caption{{\bf Identification of optimal nonlinear constitutive response via DEM simulation for a single impact condition.} (A) Feasible solutions of nonlinear spring coefficients $c_2$ and $c_3$. The black line represents $c_2=-\sqrt{3c_3}$, the red line indicates $c_2=-(1+3c_3)/2$, the blue line is $c_2=-3c_3$, and the green line is the zero strain energy throughout the whole range, $c_2=-3/2-3c_3/4$. (B) Non-dimensional force-extension relationship of the best performing nonlinear spring. (C) Ratio of maximum kinetic energy at the last particle in the nonlinear chain to the initial impact energy as a function of nonlinear spring coefficients for the impact condition of $M/M_0=0.05$ and $V/V_0=1$, with $\zeta=0.01$. The lines from (A) are overlaid, and the star marker denotes the point of best performance. Normalized kinetic energy of the (D) best performing nonlinear and (E) linear material.}
\label{m05v1_c2c3_end}
\end{figure}

Before searching for optimal nonlinear constitutive responses with our DEM, we set several bounds. First, we confine the unit cell strain to $1$ in compression, and set $c_3>0$ for simplicity. Second, we restrict our search range for nonlinear coefficients $c_2$ and $c_3$ to ensure positive strain energy throughout the entire compression range. We note that keeping the linear stiffness constant, we exclude essential nonlinearities \cite{nesterenko2013dynamics}. By examining the polynomial's properties within this range, we classify the quasi-static response into three distinct zones, ``bistability'', ``monotonic increase'', and ``local maximum'', as shown in Fig.~\ref{m05v1_c2c3_end}A. Bistability (magenta area) denotes the existence of both a local maximum and minimum other than the boundaries (the local minimum does not need to fall below zero). Monotonic increase (blue area) denotes the absence of extrema. Local maximum (green area) signifies the presence of a local maximum (no local minimum existed) within the range of the length of one unit cell. More details concerning these zones is given in Supplementary Information Note 3.

Within these bounds, we first optimize for a singular impact condition ($M/M_0=0.05$ and $V/V_0=1$), where $M_0$ is half the mass of the chain, $V_0$ is the linear sound speed, and $M$ and $V$ are the dimensional impactor mass and velocity, respectively. The material is composed of $20$ particles and $\zeta=0.01$. We vary the nonlinear coefficients of the springs $c_2$ and $c_3$, aiming to minimize the maximum kinetic energy experienced at the end of the material ($KE_{non}$) normalized by that of a linear system ($KE_{lin}$) which has all of the same properties except $c_2=c_3=0$. Given dimensional particle displacement from its rest position $x_i^\ast$, kinetic energy of the $i$th particle is defined as $m(dx_i^\ast/dt^\ast)^2/2$, where $t^\ast$ is dimensional time, such that the analogous dimensionless kinetic energy is $\dot{x_i}^2/2$. Figure~\ref{m05v1_c2c3_end}C quantifies the ratio of maximum kinetic energy at the end of the material to the initial impact energy ($TE$), as a function of the coefficients of the nonlinear spring. The best performance in Fig.~\ref{m05v1_c2c3_end}C corresponds to nondimensional spring force $f(\Delta x) =\Delta x + 5.81\Delta x^2 + 7.08\Delta x^3$, where $\Delta x$ is the spring extension and positive $f$ denotes tension. This optimal spring is plotted in Fig.~\ref{m05v1_c2c3_end}B,  and is denoted by the star marker in the inset of Fig.~\ref{m05v1_c2c3_end}C. A point of particular note about this identified optimum, is that it is not of the form one would expect based on the conventional design approach for bistable energy absorption at lower rates \cite{shan2015multistable}, where net positive energy is locked into strain energy when snapping from the undeformed state to its second stable equilibrium (\textit{i.e.}, the area under the curve from $\Delta x=0$ to the unstable equilibrium point (at $\Delta x \approx -0.2$), is less than the area under the curve from the unstable equilibrium to the second stable equilibrium point (at $\Delta x \approx -0.6$). Rather in the case of our optimum, the first area is smaller than the second. Around this point of best performance, significant performance sensitivity to the constituent nonlinear parameters can be seen (see the zoomed-in view in Fig. \ref{m05v1_c2c3_end}C). To describe this sensitivity more quantitatively, in Supplementary Information Note 4 we calculate the gradient of the inset in Fig. \ref{m05v1_c2c3_end}C, which shows $\partial(\log_{10}(KE_{non}/TE))/\partial c_2$ and $\partial(\log_{10}(KE_{non}/TE))/\partial c_3$ approaching $\pm10$. This metric thus suggests that, \textit{e.g.}, a $0.1$ difference in $c_2$ or $c_3$ can result in up to a $10\times$ change in peak $KE$. However, we emphasize that near our chosen optimum, the sensitivity is significantly lower, with $\partial(\log_{10}(KE_{non}/TE))/\partial c_2=0.32$ and $\partial(\log_{10}(KE_{non}/TE))/\partial c_3=0.07$. Spatiotemporal responses of kinetic energy of the optimal nonlinear and linear chain are shown in Fig.~ \ref{m05v1_c2c3_end}(D) and (E), respectively. In Fig.~\ref{m05v1_c2c3_end}(D) and (E), while both materials exhibit an initially sharp pulse, energy from the optimal nonlinear material is trapped around particle $15$, preventing further transmission. 

\begin{figure}[ht!]
\centering
\includegraphics[width=17.5cm]{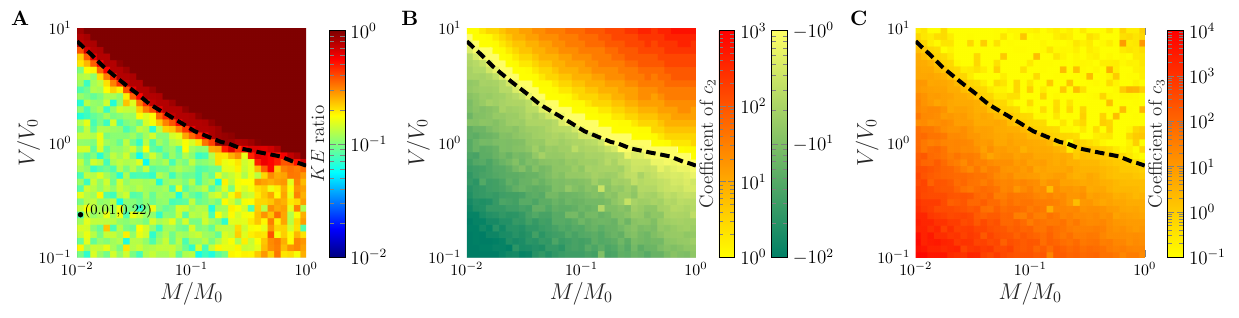}
\caption{{\bf Identification of optimal nonlinear coefficients via DEM simulation for varied impactor conditions.} (A) Optimal kinetic energy ratio ($KE$ ratio) as a function of impact conditions and corresponding nonlinear spring coefficients (B) $c_2$ and (C) $c_3$ for a material of $N=20$ and $\zeta=0.005$. The color bar in (A) is saturated at a $KE$ ratio of unity. The dashed black line denotes the onset of self contact ($\Delta x<-1$) in the linear system. The dot in (A) denotes the high performing case used for mesostructure design. }\label{mv_c2c3_ratio_end}
\end{figure}

In the subsequent analysis, we conduct an optimization wherein we look for optimal nonlinear coefficients for varied impactor mass and velocities. We use the same ratio of the maximum $KE$ of the last particle of  the nonlinear material normalized by that of the linear material as a performance metric (denoted henceforth as ``$KE$ ratio''), wherein a lower ratio indicates superior performance of the nonlinear material. The optimal (minimum) $KE$ ratios with respect to $M/M_0$ and $V/V_0$ and corresponding nonlinear spring parameters $c_2$ and $c_3$ can be seen in Fig.~\ref{mv_c2c3_ratio_end}. In contrast to Fig.~\ref{m05v1_c2c3_end}, we use lower damping ($\zeta=0.005$), chosen to emulate that of the polycarbonate springs used in our experimental realization. Additional simulation results of $KE$ ratios for increased damping and greater discreteness (more unit cells) are available in Supplementary Information Note 5 and indicate the potential for $KE ~\text{ratio} < 10^{-2}$ in the latter case. The damping value used in Fig.~\ref{mv_c2c3_ratio_end} was chosen by measuring the low-amplitude resonance of a single connector and nonlinear spring unit (see Supplementary Information Note 6). This is an important point to highlight, in that the characterizing the damping at low amplitudes allows us to temporarily discard its interplay with the nonlinear spring. This damping can be thought to stem from the intrinsic damping of the polycarbonate. Using the model of the nonlinear spring with the linear damper in parallel, thus allows us to simulate greater effective damping that may come from their dynamical interplay at higher amplitudes. Simulation cases of the nonlinear material where self-contact occurs are discarded from consideration of the optimal performance. 

There is a clear boundary where the nonlinear chain does not outperform the linear, which is correlated with the occurrence of self-contact within the linear chain (dashed black line in Fig.~\ref{mv_c2c3_ratio_end}). At impactor velocities and masses below this threshold, the nonlinear materials exhibit significantly enhanced mitigation effectiveness. As can be seen in Fig. \ref{mv_c2c3_ratio_end}A, at a relatively low to moderate level of impactor mass ($\sim0.01$ to $0.2M_0$), optimal nonlinear materials exhibit a mitigation capability ($KE$ ratio) that is over ten times better than the linear material. As the impactor mass and velocity increase and cross over the dashed black line (orange to red area near the dashed line in Fig. \ref{mv_c2c3_ratio_end}A), the priority shifts to preventing contact between unit cells, leading to a comparatively impaired energy-absorbing performance. In future studies, such self contact could be explored as a form of nonlinearity, and a design feature instead of a constraint. 

A specific optimal solution (pair of coefficient ratios) was chosen from a set of multiple solutions showing high performance $KE$ ratio. We further downselected, looking for a solution also with relatively low strain, making it more amenable to experimental implementation. The chosen solution is a nondimensional nonlinear mechanical response of the form $f(\Delta x) =\Delta x + 87\Delta x^2 + 1778\Delta x^3$ in compression, with an impact condition of $M/M_0=0.01$ and $V/V_0=0.22$ (marked in Fig.~\ref{mv_c2c3_ratio_end}A). This nonlinear constitutive law is used as the target for the mesostructure design, with the goal of finding a geometry that yields that nonlinear mechanical response.

\subsection*{Shape optimization for desired nonlinear constitutive law}

\begin{figure}[ht!]
\centering
\includegraphics[width=17.5cm]{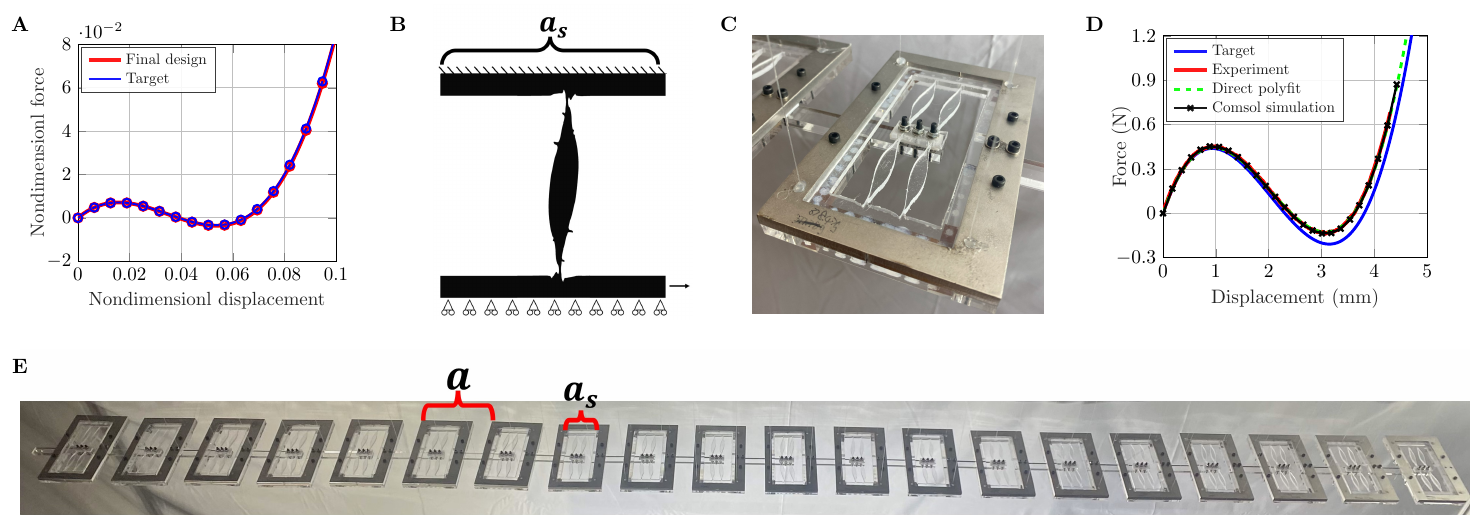}
\caption{{\bf Shape optimization to identify geometry giving DEM identified desired constitutive law, and experimental realization of the spring and chain.} A-B) Results of the shape optimization, with the target and final force-displacement curves shown in (A), and the optimized design shown in (B). C) The fabricated polycarbonate unit cell, consisting of four optimized springs (as seen in (B)) and a rigid frame to impose boundary conditions. D) The quasi-static test of the unit cell shown in (C), compared against the behavior of a single spring with perfectly imposed boundary conditions as simulated via COMSOL FEM simulation. E) The full chain of 20 unit cells, hung from a frame. The chain is clamped to the left of the leftmost unit cell, imposing a zero displacement boundary condition. The impact occurs at the right end of the chain. The unit cell and spring length scales are labeled.}\label{topo_optimization}
\end{figure}

In order to find a spring geometry that gives the desired nonlinear response, a two-dimensional shape optimization approach is taken, using a level-set optimization method (as detailed further in the Methods and Ref. \cite{QuasiOptArxivPlaceholder}). In summary, a third order polynomial is fit to the calculated force-displacement behavior of the structure, and the objective of the optimization problem is taken as the ratio of nonlinear to linear terms (\textit{e.g.}, as shown in Fig.~\ref{inverse_design_chart_flow}C and Fig.~\ref{topo_optimization}A). The exact form of the objective function is

\begin{equation}
    \min\limits_{\Omega} \sum \left( \frac{c_i}{c_j}\frac{T_j}{T_i} - 1 \right) ^2,
    \label{Eq1}
\end{equation}

\noindent where $c_i$ and $c_j$ represent the current polynomial coefficients, and $T_i$ and $T_j$ represent the target coefficients. By taking the ratio of the polynomial terms, the nonlinearity of the structure is decoupled from the linear stiffness, allowing the optimizer more design freedom, leading to more robust convergence. To further aid in the navigation of the design space, we select an initial condition for our optimization process, which displays qualitatively similar behavior to that desired (see Supplementary Information note 6).

The boundary conditions applied are depicted in Fig.~\ref{topo_optimization}B (fixed on top, roller on bottom, applied displacement on bottom 10\% of the right boundary). When realizing our nonlinear chain, we employ comparatively rigid frames around the designed spring to mimic fixed boundary conditions (as can be seen in Fig.~\ref{inverse_design_chart_flow}D and Fig.~\ref{topo_optimization}C) and rigid connectors between the springs to allow relative movement of the masses. With the addition of these components we have a unit cell length, $a$, and a nonlinear spring design domain length that encompasses only a portion of this larger unit cell. We call this portion of $a$ the spring length, and denote it by $a_s$ (and, hereafter, the subscript $s$ is used to refer to parameters defined on the scale of the spring). Figure \ref{topo_optimization}E highlights the difference between these two length scales. Because the DEM-identified polynomial constitutive law is expressed as a function of strain, and the strain experienced by the spring across $a_s$ is different than that experienced by the entire unit cell across $a$ for a fixed applied displacement, the targeted polynomial is therefore scaled accordingly. We take the ratio of nonlinear terms as $R = (c_i \epsilon^i)/(c_1 \epsilon)$ and $R_s = (c_{s,i} \epsilon_s^i)/(c_{s,1} \epsilon_s)$ on the scales of $a$ and $a_s$, respectively, with $i$ representing the order (or power) of the term and $\epsilon$ representing the maximum strain experienced on the corresponding length scale. In order for equivalent degrees of nonlinearity to be displayed at different scales, we take $R = R_s$ at the maximum strain considered in each scale, and solve for updated $c_i$ or $c_{s,i}$ terms.   

The optimization target (recall, on the unit cell scale, identified above as $f(\Delta x) = \Delta x - 87\Delta x^2 + 1778\Delta x^3$) can therefore be expressed on the spring length scale as $f(\Delta x_s) = \Delta x_s - 41.064 \Delta x_s^2 + 396.11 \Delta x_s^3$, by setting $R = R_s$ and solving for $c_{s,i}$. The final optimized structure, shown in Fig.~\ref{topo_optimization}A-B, achieved a simulation force-displacement law on the spring scale of $f(\Delta x) = \Delta x_s - 40.712 \Delta x_s^2 + 397.535 \Delta x_s^3$, and an experimental one of $f(\Delta x) = \Delta x_s - 40.654 \Delta x_s^2 + 397.274 \Delta x_s^3$ (shown in Fig. \ref{topo_optimization}D), resulting in an experimental percent difference between targeted and obtained polynomial ratios of $0.294\%$ for the third order ratio and $0.998\%$ for the second order ratio. 

As the multilevel optimization methodology is a loop, in which inputs from the DEM and shape optimization feed back into one another, the design process can be followed in either direction, which can be exploited for exploration of the space. For example, while one may begin with a desired impact condition, which is then fed into the DEM simulation followed by the mesostructure optimization, one might just as easily begin with some nonlinearity and feed this into the DEM simulation to explore what impact conditions it might perform well (or poorly) for. Similarly, one might find that the physical design domain requires constraints, or that optimized structures exhibit structural damping, which diverges significantly from predictions, and these updated considerations can be flowed back up to the DEM simulation to search for the updated optimum conditions. 

In the example shown herein, we strive for generality, thus there was no single impact condition specified. The solutions were chosen which exhibited low impactor mass and velocity, relatively smooth sensitivity to impact conditions (discussed in more detail below), and low maximum strain in order to avoid the onset of plasticity in the fabricated structures. One can just as well begin with a strict impact condition requirement, and flow through the multilevel optimization methodology accordingly. If larger mass and velocity, and therefore larger strains, are desired, the mesostructure optimization portion can be undertaken with a Neo-Hookean material model, and the real-world structures fabricated out of some soft, flexible material, such as silicone (see Ref. \cite{QuasiOptArxivPlaceholder}).    

\subsection*{Experimental validation}

In our experimental realization of the chains, the spring scale $a_s=59$ mm was chosen and incorporated into the rigid frame and connector, resulting in a unit cell length $a=125$ mm. A chain of twenty unit cells (a single unit cell is shown in Fig. \ref{topo_optimization}C) was fabricated and hung from a frame in order to minimize friction, as shown in Fig.~\ref{topo_optimization}E (see Note 7 in the Supplementary Information for a more detailed breakdown of the chain design). Impact tests were undertaken where the mass and velocity of the impactor were chosen based upon the optimal conditions identified in the DEM results, in the case with an impactor mass of $M = 40$ g and a velocity of $V = 1.37$ m/s. Data was collected through the use of several cameras positioned along the length of the chain, allowing digital image processing to be used to track the impact wave across the length of the system. In addition, a laser Doppler vibrometer (see Supplementary Information Note 7) was pointed at the last unit cell in the system, allowing for a second measurement of the velocity of the last unit cell. Impact tests were repeated several times. A similar chain of twenty linear unit cells, with similar linear stiffness and mass (the mass of the linear and nonlinear unit cells are $400.4$ g and $398.8$ g, respectively) values, was then constructed to act as a control for comparison against the nonlinear chain (see Supplementary Information Note 6 for more details), and the impact tests were repeated. It is noted that for the kinetic energy transmission ratio considered herein, the magnitude of the linear stiffness (even if different between the linear and nonlinear chain) does not matter (see Supplementary Information Note 8).

\begin{figure}[ht!]
\centering
\includegraphics[width=17.5cm]{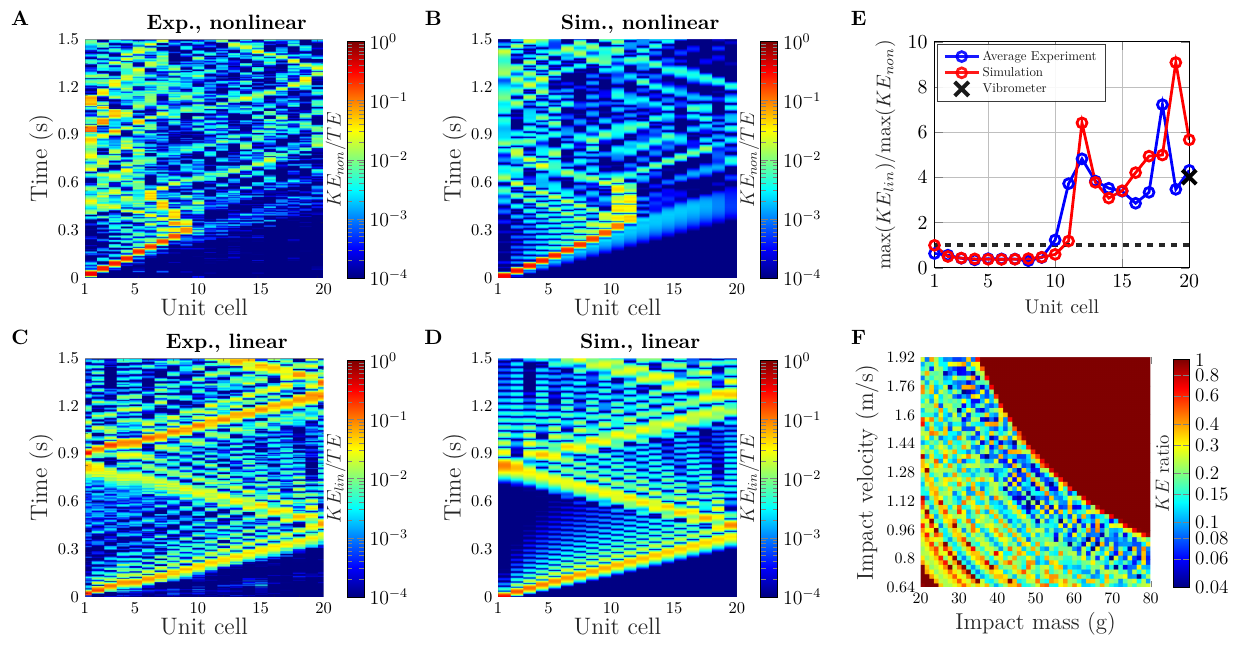}
\caption{{\bf Experimental validation of the optimal nonlinear and linear chains, compared with simulation.} A-D) Spatiotemporal evolution of kinetic energy in the system (normalized by the input kinetic energy, or initial total energy). The nonlinear chain is shown in A (experiment, nonlinear trial 1) and B (simulation), while the linear case is shown in C (experiment, linear trial 1) and D (simulation). The experimental spatiotemporal plots include a smoothing of displacement values to assist with noise induced by differentiating the discrete time camera data. The spatiotemporal plots for other trials are included in Supplementary Information Note 7. E) A ratio of maximum kinetic energy (linear/nonlinear) seen at each unit cell for both experiment and simulation. A value greater than 1 indicates superior performance of the nonlinear chain as compared to the linear. The experiment values are the average taken from three experimental trials. The X marks the average experimental value recorded by the vibrometer (see Supplementary Information Note 7 for the full data sets), which collected data from only the last unit cell. F) The simulated sensitivity of the $KE$ ratio (truncated at 1) to impact conditions, wherein the simulation was run with the coefficients that were found experimentally from the nonlinear spring in Fig.~\ref{topo_optimization}D (that is, the physically achieved coefficients). }\label{exp_compare}
\end{figure}

Several key metrics were examined to confirm the performance of the system, the results of which are summarized in Fig.~ \ref{exp_compare}. Foremost among these results is the velocity (or kinetic energy) which was transmitted to the end of the chain. Measured spatiotemporal kinetic energy responses are shown in Fig.~\ref{exp_compare}A-D, in which we can see that the nonlinear cases dissipate and trap kinetic energy through unit cell snapping, preventing much of it from reaching the right (or protected) end of the chain (a large portion of the kinetic energy is seen to remain, reflecting back and forth, in the first 9-11 unit cells in panels A and B). We note an excellent match between simulation (using the experimentally fit coefficients taken from the quasi-static force-displacement curve shown in Fig.~\ref{topo_optimization}D) and experiment in these spatiotemporal plots. Figure~\ref{exp_compare}E shows a comparison of the ratio (of linear to nonlinear cases) of the maximum kinetic energy seen at each unit cell, in both experiment (averaged across three trials) and simulation. We note that the ratio shown in Fig.~\ref{exp_compare}E is the inverse of $KE$ ratio, for ease of visualization, such that larger numbers denote better impact protection. We see a larger discrepancy at the last two particles (Fig. \ref{exp_compare}E), which we attribute to non-ideal boundary conditions.  We note that damping characterization (see Methods section) suggest the damping in both chains is similar ($\zeta$ of 0.005 for nonlinear and 0.003 for linear). Further, we show in Supplementary Information Note 6, that were the linear chain to have higher damping, \textit{e.g.}, $\zeta=0.006$, this would have negligible effect on the $KE$ ratio. This highlights the value of nonlinear wave manipulation, where superior performance can be seen without heavy reliance upon damping. 

A point of particular note, is that although the targeted conditions show excellent predicted performance, the behavior can be sensitive to small variations in impactor mass and velocity, as seen in the simulation data of Fig.~\ref{exp_compare}F. The variability of the mass-velocity space is immediately apparent, with several very small regions of excellent performance (low $KE$ ratios) surrounded by oscillating regions of lower performance (relatively higher $KE$ ratios), and even several points of poor performance ($KE$ ratio $>1$). The optimum performance region targeted in this work sought out a region with relatively low drops in performance (relative to other regions explored via simulation), but nonetheless there still exist oscillatory performance impact conditions nearby. In order to more quantitatively describe this sensitivity, in Supplementary Information Note 4, we calculate the gradient of Fig.~\ref{exp_compare}F, which shows $|\partial(\log_{10}(KE \text{ratio}))/\partial M|$ can reach near $1$ g$^{-1}$ and $|\partial(\log_{10}(KE \text{ratio}))/\partial V|$ can reach up to $40$ s/m. This means that, for the most sensitive regions of the impact conditions landscape, a change in 1 g of impactor mass can result in an up to $\sim 6.7 \times$ change in $KE$ ratio, or a 0.01 m/s change in impactor velocity can result in an up to $\sim 2.5 \times$ change in $KE$ ratio. However, as with the case of the sensitivity to the nonlinear stiffness parameters, near our chosen optimum, the sensitivity is significantly lower, with $\partial(\log_{10}(KE \text{ratio}))/\partial M=0.16$ g$^{-1}$ and $\partial(\log_{10}(KE \text{ratio}))/\partial V=-7.8$ s/m. This sensitivity is amplified due to the presence of bistability, in which the snapping (or lack thereof) of a single unit cell can push the system from one qualitative behavior well to another (see Supplementary Information Notes 4 and 9 for more in depth analysis and an example, respectively). The impactor velocities seen herein were not precise (ranging from $1.36-1.41$ m/s, see Supplementary Information Note 7), and this, coupled with the aforementioned sensitivity provides an insight into variations in chain performance---namely, that the slight variations in impactor velocity we see in experiment have the potential to easily knock the system out of its optimal performance region, into one in which poorer performance is to be expected. This phenomenon reveals important characteristics regarding the sensitivity of the system, and more generally nonlinear dynamical systems wherein bifurcation can cause sharp changes in behavior \cite{strogatz2018nonlinear}. We believe the proximity of regions of poor performance to regions of good performance (\textit{e.g.}, Fig.~\ref{exp_compare}F) motivates a consideration of nearby conditions in future work. For instance, we expect there are application scenarios in which a region of reduced sensitivity to changes in stimuli may be desirable at the expense of slightly lowered performance. Despite the sensitivity of the system, however, per Fig.~\ref{exp_compare}E the kinetic energy ratio ($KE_{lin}/KE_{non})$ remains greater than 1 (superior performance of the nonlinear chain compared to the linear) for both simulation and experiment once a critical number of unit cells has passed (unit cell 11 for the simulation, unit cell 10 in experiment). 

\section*{Discussion}

In this work, we have demonstrated a bilevel inverse design methodology for nonlinear wave tailoring in a mechanical system. Including the possibility of more highly discretized materials (\textit{e.g.}, $100$ unit cells, see Supplementary Information Note 5), our optimizations of nonlinear constitutive laws suggest the potential for over two orders of magnitude improvement (reduction) in kinetic energy transmission via the use of a nonlinear material compared to a comparative linear material. Our initial experiments on materials with fewer unit cells ($20$) show close agreement with our simulations, validating our bilevel design approach. We also observed large sensitivity of the kinetic energy transmission metric to small changes in nonlinearity and impact conditions, which we suggest highlights the need for high precision, free-form nonlinearity tailoring, as well as future study of nonlinear dynamical design for robustness-of-response to the presence of defects, disorder, and stimuli variation. Within the considered topic of impact mitigation, a near term question of future interest would be how such optimal nonlinearities would change for a different metric such as maximum transmitted force or peak tensile stress anywhere in the material. Similarly, although our proof of concept study focused on impact, we envision a wide range of applications that could make use of precisely tailored nonlinear signal transformation such as mechanical computing, acoustic signal, and image processing, vibration mitigation, and materials that autonomously respond and conduct directed work via rapid shape-change. Additional future research areas include extension of this method to less scale separated dynamical (no lumped masses), higher-dimensional, irreversible constitutive, heterogeneous material, active, and coupled-physics settings. 

\section*{Methods}

\subsection*{DEM Simulations}
We numerically integrate nondimensional Eq. (S8) in Supplementary Information Note 2 using a Runge-Kutta algorithm (ode45 in MATLAB) with impactor velocity $V/V_0$ and mass $M/m$ applied to the impactor particle. The output and maximum integration timestep is selected by estimating the highest nondimensional frequency $f_{max}=\frac{1}{2\pi}\sqrt{\frac{k_{max}}{M_0}}$, where $k_{max}=\max(1,1+2c_2+3c_3)$, such that the timestep $\Delta t=1/(1000f_{max})$. The nondimensional displacement and velocity tolerances are set to $10^{-10}$. The total energy conservation is checked for an undamped DEM system, and deviation is less than 0.1\% over the entire simulation duration. 

\subsection*{Structural Optimization}

We include here further details about the shape optimization of the spring geometry, while noting that a full description of the method can be found in our previous work \cite{QuasiOptArxivPlaceholder}. A level-set method is used to perform a nonlinear, displacement control, 2D continuum shape optimization. The spring is modeled using a 2D plane stress condition, with a Kirchhoff material model (linear elastic material model, with geometric nonlinearity included), as we have found this material model to be a good representation of polycarbonate throughout displacement ranges which do not reach the plasticity threshold of the material. The optimization performed herein was solved with a uniform mesh consisting of $750 \times 750$ quadrilateral elements. A symmetry boundary condition (rollers) is applied along the bottom edge of the optimization domain, such that only one half of the spring need be simulated. The top edge of the domain has a fixed boundary condition enforced. A 1D applied displacement is applied at the bottom right edge of the domain, where the connector to the adjacent spring would exist in the physical system. These conditions, along with the initial condition supplied to the optimizer, are illustrated in Note 6 in the Supplementary Information.

Prior to manufacture and experimental tests, the force displacement output from the optimization process was further confirmed via a FEM simulation in COMSOL Multiphysics. Following the verification process outlined in our previous work \cite{QuasiOptArxivPlaceholder}, the optimal level set was converted to a 2D geometry in COMSOL, and a quasi-static simulation was performed. This simulation similarly used 2D plane stress, with a Kirchhoff material model and the same applied boundary conditions as the optimization process. The COSMOL simulation, however, employed a body-fitted mesh consisting of 20,447 triangular elements. The results, shown in Fig. \ref{topo_optimization}D, confirm the accuracy of the optimization simulation.

As outlined in Eq.~\ref{Eq1}, the objective function for the optimization is taken to be the ratio of polynomial coefficients resulting from the force-displacement curve of the current design. This polynomial is fit to the force-displacement curve using the equation $c = V^{-1}C$, in which $V$ is the Vandermonde matrix, and $C$ is compliance (we note that herein, we use lowercase $c$ to refer to polynomial coefficients, and uppercase $C$ to refer to compliance). Fitting to compliance, rather than force, is done for ease of sensitivity calculation, as compliance optimization is a well known optimization problem. Noting that $C = F \Delta x$ (in which $\Delta x$ is displacement), it is a simple matter to transition between force and compliance. We can then write the sensitivity for the polynomial ratio objective as a combination of known compliance sensitivity terms and sensitivity terms related to the fitting of the polynomial.

\begin{equation}
    \frac{\partial c_i}{\partial \Omega} = \sum_{j=1}^{m}\frac{\partial c_i}{\partial C_j}\frac{\partial C_j}{\partial \Omega} = \sum_{j=1}^{m} V^{-1}(i,j)\frac{\partial C_j}{\partial \Omega},
    \label{Eq2}
\end{equation}

\noindent where $\Omega$ refers to the level set domain, $i$ refers to the degree of polynomial, $j$ refers to the displacement step, and $m$ is the total number of displacement points under consideration (see our previous work for more details\cite{QuasiOptArxivPlaceholder}).

\subsection*{Unit Cell Design and Manufacturing}

After the nonlinear spring had been designed via optimization, the design was resized to the experiment scale (wherein the length of one spring, $a_s$, was taken to be 59 mm). The single spring design was reflected over the symmetry condition, and in order to minimize the non-longitudinal motion of the masses, a second set of nonlinear springs was added in parallel to the first. A single unit cell (shown in Fig.~\ref{topo_optimization}C) was fabricated by cutting the shape out of a polycarbonate sheet via computer-numerical-control (CNC) milling. This milled shape was then sanded down at the edges to fit smoothly into a rectangular stainless steel frame, which was commercially manufactured via sheet cutting, and secured in-plane with an acrylic backing bolted into the frame and glued to the milled spring (see Note 7 in the Supplementary Information for a visualization). We note the design of the frame surrounding a spring is particularly important in properly imposing boundary conditions, and thus critical in matching simulation predicted force displacement curves. The choice of a stainless steel frame herein was made in order to meet these criteria, while still allowing for the milling of the spring to fit within the manufacturing constraints of the manufacturing equipment. 

A single unit cell was tested quasi-statically to confirm the performance of the spring, as shown in Fig.~\ref{topo_optimization}D. Repeated, cyclic quasi-static loading tests were also performed to confirm that no onset of plasticity or fatigue occurs during the dynamic experiment (see Supplementary Information Note 6). 

\subsection*{Data Collection}
The motion of each unit cell was captured with the use of cameras mounted above the chain. The length of the chain necessitated the use of multiple cameras to capture the motion all unit cells. Each experimental trial therefore has four videos associated with it, shot in slow motion at 240 fps on four iPhones, with each video overlapping by at least one unit cell to ensure proper spatial synchronization and the capture of all motion. For each trial, the cameras and unit cells were spaced such that camera 1 captured unit cells 1-6, camera 2 captured unit cells 4-11, camera 3 captured unit cells 9-16, and camera 4 captured unit cells 14-20. Video processing was performed in MATLAB. Temporal synchronization between each video was achieved by a visual signal which caused a fluctuation in intensity of light, which could then be detected in each video and used to synchronize the times between the recording devices. Tracking of each particle was achieved with the use of colored markers on each unit cell. Post processing in MATLAB allowed each frame to be separated into discrete color channels, allowing for the isolation and tracking of red, blue, and green markers (blue was used for the impactor, green for the back edge of each unit cell, and an additional red marker for unit cells which were covered by overlapping camera fields of view, in order to enable spatial synchronization of the data from each camera).

\subsection*{Damping Characterization}

In order to characterize the damping ratio $\zeta$ of the manufactured springs, the power spectrum was experimentally measured, generated from a low-amplitude $1$ ms duration square pulse excitation applied to a single spring via electrodynamic shaker. A Lorentzian function was numerically fit to the response for both the nonlinear and linear spring cases. The experimental setup for both cases are shown in Supplementary Information Note 6, where the single unit cell was hung horizontally using fishing lines, and the end of the connector, which was connected to the center of the spring via bolts and nuts, was clamped to impose a fixed boundary condition (FBC). The pulse excitation was provided by a function generator (FG, Tektronix AFG3022C), controlled through MATLAB, and applied via an electrodynamic shaker (The Modal Shop K2007E01) equipped with a stinger, whose tip was manually set at a small distance (within $1$ mm) away from the side surface of the unit cell. A laser Doppler vibrometer (LDV, Polytec PSV 400) was used to record the dynamical response of the unit cell in velocities. Data was collected by measuring the side surface of the unit cell (averaged three times for each test), where the LDV and the FG were synchronized through a common trigger signal for repeatability of the averaging process. Measured dynamical responses, which were processed to account for the tilted angle between the LDV scanning head and the unit cell, are shown in Supplementary Information Note 6.
The normalized power spectrum was obtained by taking the square of the Fourier transform (FFT) of time domain data and then normalizing its maximum value. 
The FFT was taken using the built-in \textit{fft} command in MATLAB. 
The normalized $|$FFT$|^2$ result was fit to a Lorentzian function of the following form 
\begin{equation}
    L(f) = \frac{A}{\pi}\frac{\Gamma/2}{(f-f_0)^2+(\Gamma/2)^2},
    \label{eq:lorentzian}
\end{equation}
where $A$ is the amplitude parameter, $f_0$ is the central frequency, and $\Gamma$ is the full width at half maximum in frequency. 
The fitting process was done by using the built-in \textit{fit} command in MATLAB. 
The $Q$ factor was found by taking $f_0/\Gamma$ from the numerical fitting, which are approximately $92$ for the nonlinear and $177$ for the linear spring unit cell.  
Based on these $Q$ factors, by using the equation $\zeta = 1/(2Q)$, we calculated the damping ratios $\zeta$ as $0.005$ and $0.003$ for nonlinear and linear chains, respectively, which were used in the numerical simulations.

\bibliography{Ref.bistable_metamateirals_optimization}

\section*{Acknowledgements}

This work was supported by the UC National Laboratory Fees Research Program of the University of California, Grant Number L22CR4520. B.M. acknowledges support from the U.S. Department of Energy (DOE) National Nuclear Security Administration (NNSA) Laboratory Graduate Residency Fellowship (LRGF) under Cooperative Agreement DE-NA0003960. I.F. acknowledges support from the Department of Defense (DoD) through the National Defense Science \& Engineering Graduate (NDSEG) Fellowship Program. 

\section*{Author contributions statement}

N.B., and H.A.K. designed research; B.M., H.X. I.F., K.Q. N.B., and H.A.K. performed research, K.Q., B.M., and I.F. conducted the experiments, B.M., H.X. I.F., K.Q. analysed the results; and B.M., H.X. I.F., K.Q. N.B., and H.A.K. wrote the paper. 

\section*{Data Availability}
Two sets of four videos (reduced resolution and sped up via downsampling by $8\times$ to meet file size limitations) corresponding to the data shown in Fig.~\ref{exp_compare}A and Fig.~\ref{exp_compare}C are included as part of the Supplementary Information. The same naming convention is used as in Supplementary Information Table 1, but with the suffix ``\_SpedUpLowRes''. 

All related videos regarding the data in the main text are uploaded to: Boechler, Nicholas (2024), “Customizable wave tailoring materials enabled by nonlinear bilevel inverse design 1”, Mendeley Data, V1, doi: 10.17632/2wgwfy2wfg.1 

Videos related to the data in Supplementary Information Note 9 are uploaded to: Boechler, Nicholas (2024), “Customizable wave tailoring materials enabled by nonlinear bilevel inverse design 2”, Mendeley Data, V2, doi: 10.17632/6bg6hr5kyr.2

\section*{Competing Interests}
The authors declare no competing interests.

\end{document}